\documentstyle{mn}

\input epsf

\begin{document}
\journal{Sussex preprint SUSSEX-AST 95/12-1, astro-ph/9512102}
\title[CDM with a cosmological constant]{Cold dark matter models with a 
cosmological constant}
\author[A.~R.~Liddle et al.]{Andrew R.~Liddle,$^1$ David H.~Lyth,$^2$ Pedro 
T.~P.~Viana$^1$ and Martin White$^3$\\
$^1$Astronomy Centre, University of Sussex, Falmer, Brighton BN1 9QH\\
$^2$School of Physics and Materials, University of Lancaster, Lancaster LA1 
4YB\\
$^3$Enrico Fermi Institute, University of Chicago, 5640 S.~Ellis Ave, 
Chicago, Illinois 60637,~~~USA}
\date{Accepted 1996 April 29. Received 1996 April 19; in original form
December 18}
\maketitle
\begin{abstract}
We use linear and quasi-linear perturbation theory to analyse cold dark
matter models of structure formation in spatially flat models with a
cosmological constant. Both a tilted spectrum of density perturbations and a 
significant gravitational wave contribution to the microwave anisotropy are 
allowed as possibilities. We provide normalizations of the models to 
microwave anisotropies, as given by the four-year {\it COBE} observations, 
and show how all the normalization information for such models, including 
tilt, can be condensed into a single fitting function which is independent 
of the value of the Hubble parameter. We then discuss a wide variety of 
other types of observations. We find that a very wide parameter space is 
available for these models, provided $\Omega_0$ is greater than about 0.3, 
and that large-scale structure observations show no preference for any 
particular value of $\Omega_0$ in the range 0.3 to 1.
\end{abstract}
\begin{keywords}
cosmology: theory -- dark matter.
\end{keywords}

\section{Introduction}

Even before the {\it COBE} satellite's detection of microwave background 
anisotropies, a popular strategy for dealing with the perceived ills of the 
standard cold dark matter (CDM) model of structure formation was to reduce 
the matter density. This shifts the redshift of matter--radiation equality, 
distorting the shape of the spectrum of density perturbations to provide a 
good fit to the galaxy correlation function. In order to retain 
compatibility with typical models of inflation, a flat universe is desired 
which can be brought about by adding a cosmological constant $\Lambda$ to 
the model (Peebles 1984; Turner, Steigman \& Krauss 1984; Efstathiou, 
Sutherland \& Maddox 1990; Kofman, Gnedin \& Bahcall 1993; Stompor, 
G\'{o}rski \& Banday 1995; Klypin, Primack \& Holtzman 1995). We will 
denote such models $\Lambda$CDM models.

The $\Lambda$CDM models are most commonly considered in the context of a 
scale-invariant (Harrison--Zel'dovich) primordial spectrum of density 
perturbations \cite{ESM,SGB,KPH}, corresponding to spectral index $n=1$.
Another limitation usually placed on the models is that the gravitational 
waves should not contribute significantly to the large-scale microwave 
background anisotropy. However, within a given model of inflation both the 
spectral index and the gravitational wave contribution are determined, and 
while many models do give $n$ very close to $1$ and negligible gravitational 
waves, there are other models that do not. It is therefore realistic to 
allow both $n$ and the gravitational wave contribution to vary when 
confronting a model with the data, and we will adopt that viewpoint in this 
paper. Observational constraints on $n$ and the gravitational wave 
contribution will in the future place powerful constraints on models of 
inflation, and hence on the nature of the fundamental interactions at very 
high-energy scales. Kofman et al. \shortcite{KGB} have previously considered 
one example of the model with (mild) tilt and a gravitational wave 
contribution. Because the $\Lambda$CDM model has generated further interest 
recently \cite{KT,OS,BPN}, we believe that the full parameter space merits 
detailed study, and in this paper we compare a variety of predictions of the 
model, accessible via linear and quasi-linear perturbation theory, with 
observations of various types.

We thus permit arbitrary variation of three main parameters, namely the 
matter density $\Omega_0$ expressed as a fraction of the critical 
density (the assumption always being that a cosmological constant restores 
spatial flatness), the present value of the Hubble parameter $H_0 = 100 \, h 
\; {\rm km \, s}^{-1} \, {\rm Mpc}^{-1}$, and the spectral index $n$ of the 
primordial spectrum. In addition, we will consider two possibilities 
concerning gravitational waves; first that they are negligible, and 
secondly that they have an amplitude appropriate to the power-law inflation 
model.

One of the most important observations is that of cosmic microwave 
background (CMB) anisotropies by the {\it COBE} satellite, and we provide 
careful normalizations of the models to the four-year {\it COBE} data. We 
quote a new fitting function which gives accurate normalizations for any 
combination of parameters. We also provide some comparison with 
intermediate-scale anisotropies, where the observational situation is less 
clear. We then go on to address a variety of constraints concerning 
large-scale structure, extending techniques we have already employed for 
critical-density models \cite{LLSSV} and for open universe CDM models 
\cite{LLRV}. The most closely related paper is Stompor et al. 
\shortcite{SGB}, who provided extensive comparison with {\it COBE} for the 
case of a scale-invariant ($n=1$) initial power spectrum, along with 
discussion of other large-scale structure constraints. They did not, 
however, discuss tilt or gravitational waves; nor did they attempt to 
delineate the allowed parameter region in the $\Omega_0$--$h$ plane.

\section{The Power Spectrum}

We specify the power spectrum of the density contrast, following the 
notation of Liddle \& Lyth \shortcite{LL} and Liddle et al. 
\shortcite{LLRV}, as
\begin{equation}
\label{powerspec}
{\cal P}_{\delta}(k) = \left( \frac{k}{aH} \right)^4 \delta_{{\rm H}}^2(k) 
	\, T^2(k) \, \frac{g^2(\Omega)}{g^2(\Omega_0)} \,,
\end{equation}
where $k$ is the wave number, $a$ is the scale factor, $H = \dot{a}/a$ is 
the Hubble parameter and $\Omega$ is the density parameter. Subscript `0' 
indicates the present value. One can think of the remaining terms in three 
parts. First, $\delta_{{\rm H}}^2(k)$ gives the primordial spectrum of 
anisotropies, before they are affected by any evolution. The shape is 
determined by whichever mechanism is used to create the spectrum, which we 
will assume is cosmological inflation. If $\delta_{{\rm H}}^2(k)$ is a 
constant then one has the Harrison--Zel'dovich spectrum, while $\delta_{{\rm 
H}}^2(k) \propto k^{n-1}$ corresponds to a `tilted' spectrum with spectral 
index $n$. Such a spectrum is the generic prediction of slow-roll inflation 
\cite{LL92}. The amplitude of $\delta_{{\rm H}}^2(k)$ fixes the present-day 
normalization of the spectrum. The transfer function $T(k)$ contains the 
evolution of the spectrum from its primordial form to its present form, and 
depends on cosmological parameters and the nature of the matter in the 
universe. It approaches unity on large scales (which retain their primordial 
form), and in a CDM model is time-independent well after matter--radiation 
equality. Finally, there is a factor which specifies the time, or redshift, 
dependence of the amplitude of the spectrum at late epochs. If $\Omega_0$ 
equals 1 this is carried in the $(aH)^4$ prefactor; for low-density 
cosmologies there is an additional suppression factor which we parametrize 
via a quantity $g(\Omega)$, defined below.

A quantity related to the power spectrum is the dispersion $\sigma(R)$ of 
the density field smoothed on a (comoving) scale $R$, defined by 
\begin{equation}
\sigma^2(R) = \int_0^\infty W^2(kR) \, {\cal P}_{\delta}(k) \frac{{{\rm
	d}}k}{k} \,.
\end{equation}
To carry out the smoothing, we will always use a top-hat window function 
$W(kR)$ defined by
\begin{equation}
\label{tophat}
W(kR) = 3 \left( \frac{\sin(kR)}{(kR)^3} - \frac{\cos(kR)}{(kR)^2} \right) 
	\,.
\end{equation}

For CDM models as considered here, the transfer function is accurately given 
by Bardeen et al. \shortcite{BBKS} as
\begin{eqnarray}
T_{{\rm CDM}}(q) & = & \frac{\ln \left(1+2.34q \right)}{2.34q} \times  
	\\ \nonumber 
& & \hspace*{-1.5cm}
	\left[1+3.89q+(16.1q)^2+(5.46q)^3+(6.71q)^4\right]^{-1/4} \,,
\end{eqnarray}
with $q = k/h\Gamma$, where the `shape parameter' $\Gamma$ is defined as 
\cite{SUG95}
\begin{equation}
\Gamma = \Omega_0 h \exp (-\Omega_{{\rm B}}-\Omega_{{\rm B}}/\Omega_0) \,.
\end{equation}
Here $\Omega_{{\rm B}}$ is the baryon density, which we take to equal $0.016 
h^{-2}$ as suggested by recent analyses of nucleosynthesis (Copi, Schramm \& 
Turner 1995a,b). This fit to $\Gamma$ is good for both flat and open 
universes (unless $\Omega_0$ is very low). 

We have accounted for the redshift dependence of the amplitude of the power 
spectrum by introducing a growth suppression factor $g(\Omega)$, following 
Carroll, Press \& Turner \shortcite{CPT}. This gives the total suppression 
of growth of the dispersion $\sigma(R)$ relative to that of a 
critical-density universe, and is accurately parametrized by
\begin{equation}
\label{supp}
g(\Omega) = \frac{5}{2} \Omega \left[ \frac{1}{70} +
	\frac{209\Omega}{140} -\frac{\Omega^2}{140} + \Omega^{4/7}
	\right]^{-1} \,.
\end{equation}
This formula can be applied at any value of $\Omega$. For a matter-dominated 
flat universe, the redshift dependence of $\Omega$ is given by
\begin{equation}
\label{omz}
\Omega(z) = \Omega_0 \, \frac{(1+z)^3}{1-\Omega_0 +(1+z)^3 \Omega_0} \,.
\end{equation}
Since the growth law in a critical-density universe is $\sigma(z) \propto 
(1+z)^{-1}$ (carried by the $(aH)^4$ term in equation (\ref{powerspec})), 
the redshift dependence for arbitrary $\Omega_0$ is therefore
\begin{equation}
\label{growth}
\sigma(R,z) = \sigma(R,0) \, \frac{g(\Omega(z))}{g(\Omega_0)} \, 
\frac{1}{1+z} \,.
\end{equation}

\section{Microwave Background Anisotropies}

\subsection{{\it COBE}}

With the detection of large-angle temperature fluctuations in the
cosmic microwave background, the {\it COBE} satellite made possible
for the first time an accurate normalization of models of structure
formation. At the large scales probed the spectrum can be normalized
in a regime where the theory is well understood, circumventing
complications introduced by processing of the primordial spectrum and
the relationship between observed structure and the underlying density
field.

The first-year {\it COBE} data were of low signal-to-noise ratio, with
the rms fluctuation having a 30 per cent error.  Fits to the full data
set, the angular correlation function or the rms fluctuation all gave
consistent values for the quadrupole normalization: $Q_{\rm
rms-PS}=17\pm5 \, \mu$K \cite{COBE,SB,SV,Wetal}.
 
The second year of data \cite{B94} resulted in a dramatic improvement
of the signal-to-noise ratio and a consequent increase in the degree of 
refinement of the analyses.  The two-year data constrain the large-scale
normalization to within $10$ per cent (at 1$\sigma$), with 5--7 per cent of 
this being due to irremovable cosmic and sample variance. Ironically, along 
with the better signal-to-noise ratio came ambiguity in the number to use 
for normalization (amounting to a 30 per cent discrepancy!), due in large 
part to a low quadrupole in the two-year map (Banday et al. 1994; Bunn, 
Scott \& White 1995). Normalization of models directly to the temperature 
maps became essential to obtain all the information available from the {\it 
COBE} data \cite{Getal,Bond,Bunn}. In addition, highly accurate calculations 
of theoretical predictions and their relation to large-scale structure were 
included in the analyses \cite{BSW,BS,HBS,GRSB,TB,WB,SGB,YB,Cetal,WS}, 
making the {\it COBE} normalization one of the most accurately known pieces 
of information about large-scale structure.

With the release of the four-year anisotropy maps 
\cite{Ben96,Ban96,G96,H96}, we have in hand all of the knowledge about large 
angular scale anisotropies that we can expect to obtain in the near future.
The full data set prefers a slightly lower normalization ($\sim1\sigma$) 
than the two-year data, due in equal parts to a statistical downward 
fluctuation and an improved galaxy cut \cite{G96}. Models with near 
scale-invariant spectra, such as the $\Lambda$CDM models considered here, 
are now a better fit to the data and the actual quadrupole on the sky is no 
longer anomalously low. However, while the `final' normalization is lower 
than that of the two-year data, the central value is still higher than that 
which would be obtained from considering only the rms fluctuation in the 
map.  The {\it COBE} data still cannot be summarized by one number and fits 
to the full data set are necessary to obtain a precise normalization.

For low-density CDM models, the {\it COBE} data play a vital role in 
breaking the degeneracy between flat $\Lambda$CDM models and open CDM 
models. While arguments based on structure formation alone are insensitive 
to a cosmological constant \cite{MW,M91,Lahav}, the {\it COBE} 
normalizations of the two theories are very different. We will say more on 
this in the conclusions.

Inflation predicts not only a spectrum of density perturbations, but also 
one of gravitational waves (also known as tensor perturbations). Those with 
a wavelength comparable to the size of the observable Universe can induce 
extra microwave background anisotropies over and above those caused by the 
density perturbations. In absolute generality, the contribution of these to 
the {\it COBE} normalization is independent of the spectral index of density 
perturbations, permitting an arbitrary reduction in the normalization of the 
density perturbations. For $n <1$, we choose to examine two cases. The first 
case has negligible gravitational waves, which is a prediction of many 
inflationary models even when they give significant tilt (the reason being 
that the gravitational waves are negligible unless the potential during 
inflation is within a few orders of magnitude of the Planck scale). In the 
second, the gravitational wave amplitude is the one given by power-law 
inflation (PLI); this happens to be a good approximation for the known 
models of inflation that give a significant gravitational wave contribution. 
We will also study $n>1$, but in that case we will not consider 
gravitational waves since it is hard to arrange gravitational waves of 
significant amplitude when $n$ is greater than 1.

For our assumed inflationary theories the density perturbation and
gravitational wave spectra are given by perfect power laws. We compute the 
corresponding radiation power spectrum by numerically integrating the 
coupled Einstein, fluid and Boltzmann equations to the present as discussed 
in Hu et al. \shortcite{HSSW}. We carry out a fit to the four-year {\it 
COBE} data using the method of White \& Bunn \shortcite{WB} with the 
customized galactic cut of Kogut et al. \shortcite{K96}. A full description 
of this calculation will be given by Bunn \& White (in preparation).

Defining the spectrum in the manner of equation (\ref{powerspec}) is
extremely useful, because when expressed in terms of $\delta_{{\rm H}}$ 
the normalization is {\em independent} of the present Hubble parameter 
$h$ to an excellent approximation.  Further, the scale of the {\it COBE} 
observation is large enough that details of the material content of the 
Universe are unimportant, making the normalization independent of the baryon 
density and whether or not the dark matter is multi-component. So the 
present-day normalization of the power spectrum depends only on the density 
parameter $\Omega_0$, the tilt of the spectrum $n$ and, if included, the 
amplitude of any gravitational waves that might influence the {\it COBE} 
result.

In the scale-invariant case $\delta_{{\rm H}}(k)$ is constant, and can
be specified on any scale. More generally, it is necessary to specify
the scale on which it is quoted, via
\begin{equation}
\delta_{{\rm H}}^2(k) = \delta_{{\rm H}}^2(k_0) \left(\frac{k}{k_0}
	\right)^{n-1} \,.
\end{equation}
We will choose $k_0 = a_0 H_0$, the present Hubble scale, and write 
$\delta_{{\rm H}} \equiv \delta_{{\rm H}}(k_0)$. 

We find that we are able to represent the normalization of tilted
$\Lambda$CDM models by a single fitting function
\begin{equation}
\label{norm}
\delta_{{\rm H}}(n,\Omega_0) = 1.94 \times 10^{-5} \; 
	\Omega_0^{-0.785-0.05\ln\Omega_0}
	\exp \left[ f(n) \right] \,,
\end{equation}
where 
\begin{eqnarray}
f(n) & = & -0.95(n-1)-0.169(n-1)^2 \quad \mbox{No tensors} \,;\\
 & = & \; 1.00(n-1)+1.97(n-1)^2 \quad \quad \mbox{PLI} \,.
\end{eqnarray}
These fits are accurate to 3 per cent in the range $0.2 < \Omega_0
\leq 1$ and $0.70< n < 1.2$ (the latter formula though only applying for $n 
\leq 1$). The statistical uncertainty in all of these numbers is 7 per cent, 
and there is an additional 3 per cent `systematic' uncertainty from the 
choice of galaxy cut; we combine these to obtain a 15 per cent uncertainty 
at 2$\sigma$. The fits are accurate for any reasonable $h$ and $\Omega_{{\rm 
B}}$, and for any choice of dark matter content. This extremely simple form 
is much more easily applied than the tabular data of White \& Bunn 
\shortcite{WB} and Stompor et al. \shortcite{SGB}, as well as applying to a 
much wider parameter space of models. 

Although we normalize all models to the central {\it COBE} value, we allow 
for its uncertainty by adding it in quadrature to the relative error of 
other observations constraining the amplitude.

\subsection{Intermediate-scale anisotropies}

Potentially the most powerful test structure formation models will have to 
surpass in the future is the shape of the microwave anisotropy power 
spectrum on smaller angular scales than those sampled by {\it COBE}.
It is expected that this can be measured to great accuracy by proposed
satellite observatories.  At present the observational situation is rather
uncertain, and so it is best to focus on a single quantity --- the height of
the peak at degree scales relative to the large angular scale plateau.

In a $\Lambda$CDM model, the first peak in the spectrum is located at around
$\ell \simeq 220$, with a small dependence on $\Omega_0$ and $h$.
Here $\ell$ is the multipole number in an expansion of the CMB temperature
field: $\Delta T=\sum_{\ell m} a_{\ell m}Y_{\ell m}(\theta,\phi)$ with
$Y_{\ell m}(\theta,\phi)$ the usual spherical harmonics.
The {\it COBE} data sample multipoles $\ell=2$--30, with $\ell\simeq10$
being the pivot point with changing $n$. We define a quantity
\begin{equation}
D(\ell) = {\ell (\ell+1) C_{\ell} \over 110 C_{10}} \,,
\end{equation}
where the $C_\ell$ are (as usual) the expected values of the coefficients of 
the multipole expansion: $C_\ell=\left\langle \left| a_{\ell m} 
\right|^2\right\rangle$. Focusing on $D(220)$, the predicted value is given 
by the following fitting function \cite{W}:
\begin{equation}
D(220) \simeq 5.1 \left( {220\over 10} \right)^\nu \,,
\end{equation}
where we have formed the degenerate combination of parameters 
\begin{eqnarray}
\label{eqn:nudef}
\nu &\equiv& (n-1) - 0.32 \ln(1+0.76r) + 6.8 (\Omega_{{\rm B}} h^2 - 
	0.016 ) \nonumber \\
 &&     - 0.37 \ln(2h) - 0.16 \ln(\Omega_0) - 0.65 \tau \,.
\end{eqnarray}
Here $r$ is the tensor to scalar ratio, normalized to $r = 7(1-n)$ for 
power-law inflation in the $\Omega_0=1$ and $n \rightarrow 1$ limit as in
Davis et al. \shortcite{Davetal}. This form accurately describes the peak 
height for $\Omega_0>0.3$ and $0.01 \le \Omega_{{\rm B}} h^2 \le 0.02$. For 
models with lower $\Omega_0$ the shape of the CMB spectrum is sufficiently 
different from when $\Omega_0=1$ that a case by case comparison is 
warranted.

The term depending on $\tau$ shows the reduction in the peak height due
to reionization between $z=0$ and the last scattering surface.
The optical depth to Thomson scattering for ionized fraction $x_{{\rm e}}$ 
from $z=0$ to the reionization redshift $z_{{\rm R}}$ is
\begin{equation}
\tau=0.035\, \frac{\Omega_{{\rm B}}}{\Omega_0} \, h \, x_{{\rm e}}
     \left[ \sqrt{\Omega_0(1+z_{{\rm R}})^3+1-\Omega_0}-1 \right] \,.
\end{equation}
The uncertainties in the epoch of reionization and value of $\Omega_{{\rm 
B}} h^2$ are the biggest barrier to using the measured height of the peak to 
constrain the spectral slope.

Present observations, focusing on the {\it COBE} data and intermediate-scale 
experiments which probe only around the peak\footnote{We include the 
new Saskatoon-95 data \cite{SK95} in this analysis, marginalizing over the 
14 per cent calibration uncertainty. We also include the {\it COBE}, Python 
III 3-point, SP94, MAX, and MSAM three-point data which probe predominantly 
around the first peak; see White \shortcite{W} for a discussion of these 
experiments. We do not perform any foreground subtraction in this
analysis.}, provide the constraint $\nu = -0.02 \pm 0.12$ ($\pm 2\sigma$ 
range). While this result provides a strong constraint in the case of the
$\Omega_0=1$ models, limiting the degree to which tilt can be used to reduce
small-scale power, it is not very constraining for $\Lambda$CDM models.
The reason for this is that $\Lambda$CDM models have a larger amount of 
power on degree scales than the critical models, giving them more room for a 
tilt. The upper limit on the height is compromised by the possibility of 
reionization in these models, which reduces the degree-scale power and is 
exponentially sensitive to the assumed redshift of reionization. We will 
only use the lower limit, and make the conservative assumption of no
reionization. 

We have fixed the baryon density in equation (\ref{eqn:nudef}) to the 
central nucleosynthesis value. We include the uncertainty in $\Omega_{{\rm 
B}}h^2$ by adding it in quadrature to the observational limits on $\nu$, 
though this only increases the error bar on $\nu$ to $\pm 0.13$. This 
uncertainty would become greater, however, if one were to allow 
$\Omega_{{\rm B}}h^2$ to drift outside the nucleosynthesis range. We note in 
passing that the next generation of smaller angular scale (mostly 
interferometer) CMB experiments will allow tighter constraints, and more 
importantly will suffer less from uncertainties in cosmological modelling.

\section{Large-scale structure observations}

Our use of observational data has been described extensively for the case of 
a critical-density universe in Liddle et al. \shortcite{LLSSV}, and we will 
not reproduce the analyses here except where important differences arise 
when the cosmological constant is introduced. Instead we provide a short 
summary.

\subsection{Galaxy correlations}

We use results from the compilation of data carried out by Peacock \& Dodds 
\shortcite{PD}. They supply values of the power spectrum at a selection of 
different $k$, which can be used directly to constrain the shape parameter 
$\Gamma$. In their paper they quoted a central value of $\Gamma = 0.255 + 
0.32(1/n-1)$; however, they have noted since \cite{P96,PD2} that both 
non-linear bias and non-linear evolution of the power spectrum may have 
some effect on short scales. We choose to drop the four shortest scale 
points and refit. We find $\Gamma = 0.230 + 0.28(1/n-1)$, where the 
uncertainty is plus 18 per cent and minus 15 per cent at 95 per cent 
confidence. The galaxy correlation function data are not useful for 
constraining the amplitude of the power spectrum.

\subsection{POTENT}

The bulk velocity flow has long been suggested as one of the main worries 
for the low-density model, which predicts lower velocities than its 
critical-density counterpart for the same size of density contrast. However, 
the {\it COBE} normalization for low densities is actually much higher, as 
we have seen, and it will turn out that the velocities give no useful 
constraint.
In fact, at present the bulk velocity constraint is the weakest constraint 
of all. Additional ways of using the velocity data, which permit stronger 
constraints not directly interpretable in linear theory, will be discussed 
in Section \ref{POT2}.

We use the measurement of the bulk flow around our position by the POTENT 
technique \cite{BD}, using the Mark III data set \cite{Dekel}. This contains 
the evaluation of the bulk flow in spheres about our position for a variety 
of radii. However, the bulk flow is sensitive to long wavelengths to a much 
greater extent than the dispersion of the density field, and consequently in 
any given realization these data are highly correlated. Because of this, we 
concentrate on the measurement on a single scale, $40 \, h^{-1}$ Mpc. This 
is obtained observationally first by carrying out a smoothing with a $12 \,
h^{-1}$ Mpc Gaussian to generate a continuous velocity field, and then 
velocity reconstruction is carried out to give the flow corresponding to a 
top-hat smoothing. The theoretical prediction for the rms bulk flow is 
therefore
\begin{equation}
\sigma_v^2(R) = H_0^2 f^2(\Omega_0) \! \int_0^\infty \! \! \! W^2(kR) 
	\; {{\rm e}}^{-(12\, h^{-1} k)^2} \, \frac{{\cal P}_0}{k^2} 
	\frac{{{\rm d}}k}{k} ,
\end{equation}
where $W(kR)$ is the top-hat window given by equation (\ref{tophat})
and the factor $f^2(\Omega_0)$ gives the velocity suppression. Often 
$f(\Omega_0)$ is approximated as $\Omega_0^{0.6}$, which turns out to work 
well for both open and flat models \cite{Lahav}, but for the sake of 
accuracy we numerically obtain the precise result, following Carroll et al. 
\shortcite{CPT}.

The Mark III POTENT analysis gives the bulk flow in a $40 \, h^{-1}$ Mpc 
sphere as \cite{Dekel}
\begin{equation}
\label{bf}
v_{{\rm obs}}(40 \, h^{-1} {\rm Mpc}) = 373 \pm 50 \; {\rm km} \, 
	{\rm s}^{-1}\,.
\end{equation}
The quoted error represents different strategies for coping with sampling 
gradient bias and is to be considered as an estimate of the systematic 
error. In addition there is a 1$\sigma$ random error of 15 per cent. 
However, both of these are dominated by cosmic variance, which arises 
because we have only a single measurement which is to be drawn from an 
ensemble following a $\chi_3^2$ distribution. Modelling this along with the 
observational errors as in Liddle et al. \shortcite{LLSSV}, we find that the 
95 per cent confidence limits on the estimate of the dispersion are $+295$ 
per cent and $-47$ per cent. The huge asymmetry is caused by the asymmetry 
of the $\chi_3^2$ distribution, and only the lower limit is of interest to 
us.

\subsection{Cluster abundance}

In a recent paper, Viana \& Liddle \shortcite{VL} re-analysed the constraint 
on the power spectrum from the abundance of large galaxy clusters, using 
X-ray observations and a calculational technique based on Press--Schechter 
\shortcite{PS} theory. Traditionally, the cluster abundance constraint is 
quoted on the scale $8 \, h^{-1}$ Mpc, which in linear theory corresponds 
more 
or less to the appropriate mass of a cluster. The result found was
\begin{equation}
\sigma_8 = 0.60 \; \Omega_0^{-C(\Omega_0)}  \,,
\end{equation}
where the fitting function 
\begin{equation}
C(\Omega_0) = 0.59 - 0.16 \, \Omega_0 + 0.06 \, \Omega_0^2 
\end{equation}
parametrizes the changing power-law index of the $\Omega_0$ dependence. At 
$\Omega_0 = 1$, the uncertainty was found to be $+32$ per cent and $-24$ per 
cent. As $\Omega_0$ is decreased, the uncertainty becomes slightly larger; 
this increase can be expressed by multiplying the uncertainties for 
$\Omega_0 = 1$ by a factor $\Omega_0^{0.26 \log_{10} \Omega_0}$ (though the 
difference does not really become important until $\Omega_0 < 0.3$).

\subsection{Early object formation}

For critical-density models of structure formation, such as models with both 
cold and hot dark matter, an important constraint is whether or not 
structure can form sufficiently early. The strongest constraints available 
at the moment come from the amount of gas in damped Lyman alpha systems, 
which for those models imposes new constraints on parameter regions not 
excluded by other data.

For low-density models, whether open or spatially flat, these constraints 
are much less of a concern, for the reason that structure has been growing 
much more slowly recently, and hence one automatically finds that at high 
redshift structure should be more advanced than in the critical-density 
case. Nevertheless, for completeness we perform here a calculation of the 
constraint arising from damped Lyman alpha systems, following the strategy 
based on Press--Schechter theory utilized by Liddle et al. \shortcite{LLRV} 
to extend the usual calculation to the open universe case.

The contribution of the baryons present in damped Lyman alpha systems to 
the mean cosmological mass density at some redshift $z$ is given by  
(e.g. Padmanabhan 1993, p. 347)
\begin{equation}
\rho_{{\rm DLAS}}(z) = \mu m_{{\rm p}} \,  \langle \bar{N} \rangle \, 
	\frac{{\rm d}N}{c{\rm d}t} \,,
\end{equation}
where $\mu$ is the baryonic mean molecular weight, $m_{{\rm p}}$ is the 
proton mass, $\langle \bar{N} \rangle$ is the mean H$\,${\sc I} column 
density as a 
function of redshift and ${\rm d}N/c{\rm d}t$ is the number density of
damped Lyman alpha systems per unit length interval at that redshift. As
${\rm d}N/{\rm d}z$ is a measurable quantity, the only dependence of 
$\rho_{{\rm DLAS}}(z)$ on the assumed cosmology comes from
${\rm d}z/c{\rm d}t$. Using the Friedmann equation it is straightforward to 
show that, for the cosmological models we are interested in, the amount of 
baryonic matter deduced from observation at a given $z$ is given by the 
following function of $\Omega_0 \,$:
\begin{equation}
\Omega_{{\rm DLAS}}(\Omega_{0},z) =  \Omega_{{\rm DLAS}}(\Omega_{0}=1,z) \, 
	\sqrt{\frac{\Omega_{0}}{\Omega(z)}} \;.
\end{equation}
For our cosmology, $\Omega(z)$ is given by equation (\ref{omz}),
though the above result applies even to non-flat cosmologies with only 
matter plus a cosmological constant provided that the appropriate 
$\Omega(z)$ is 
used. Expressing $\Omega_{{\rm DLAS}}(\Omega_{0},z)$ in terms of its value 
assuming a critical-density cosmology is useful since most quoted 
observational results for $\Omega_{{\rm DLAS}}(z)$ assume such a cosmology.

Estimates of the abundance of baryons in damped Lyman alpha systems have 
recently been provided at redshifts 3 and 4 by Storrie-Lombardi et al. 
\shortcite{storrie}, the redshift 3 data giving a somewhat lower result than 
the much used earlier data of Lanzetta, Wolfe \& Turnshek \shortcite{LWT}. 
The constraints from the redshift 3 and 4 points are very similar, and we 
will use the latter. Using the above expression we then have 
\begin{equation}
\label{omDLAS}
\Omega_{{\rm DLAS}}(z=4) = \left( 0.0011 \pm 0.0002 \right) h^{-1} \, 
	\sqrt{\frac{\Omega_{0}}{\Omega(z=4)}} \;,
\end{equation}
where the uncertainty is 1$\sigma$. As mentioned before, $\Omega(z)$ is 
given by equation (\ref{omz}) and at redshift 4 is very close to unity for 
any reasonable $\Omega_0$. Expression (\ref{omDLAS}) conservatively assumes 
all the gas to be in the neutral state. 

Since we assume all the dark matter to be cold, a reasonable hypothesis is 
that the total density of the systems is larger than the baryonic density by 
a factor $\Omega_0/\Omega_{{\rm B}}$. Then the fraction of mass which is in 
bound objects of mass at least the typical mass $M$ is given by
\begin{equation}
f(>M,z=4) > \left( 0.069 \pm 0.021 \right) h\, 
	\sqrt{\frac{\Omega_{0}}{\Omega(z=4)}} \;,
\end{equation}
where we have interpreted the error on $\Omega_{{\rm B}} h^2$ in Copi et al. 
\shortcite{CST2} as 95 per cent confidence and added it in quadrature. Using 
Press--Schechter theory \cite{PS}, the theoretical prediction for this 
quantity is
\begin{equation}
f(>M,z) \equiv \frac{\Omega(>M(R),z)}{\Omega(z)} = {\rm erfc} \left(
	\frac{\delta_{{\rm c}}}{\sqrt{2} \, \sigma(R,z)} \right) \,,
\end{equation}
where $\delta_{{\rm c}}$ is a threshold parameter to be fixed via $N$-body 
simulations and the smoothing to obtain $\sigma(R)$ is carried out via the 
top-hat window of equation (\ref{tophat}). 

The most conservative assumption is that the damped Lyman alpha systems have 
not fully collapsed along all their axes. This corresponds to a lower choice 
of threshold than the usual 1.7, so, following Liddle et al. 
\shortcite{LLSSV}, we adopt $\delta_{{\rm c}} = 1.5$ \cite{Mon}. The minimum 
mass of the damped Lyman alpha systems (from the assumption that they 
eventually give rise to rotationally supported discs as indicated by lower 
redshift observations \cite{LWT}) is taken as $10^{10} h^{-1} {{\rm 
M}}_{\sun}$ \cite{H95}.

\section{Miscellaneous constraints}

In this section we consider other constraints which motivate the
preferred choice of $\Omega_0$. They either fall outside the description
of large-scale structure, or correspond to constraints from large-scale
structure that we have insufficient technology to compute ourselves and
must take at face value from other papers.

\subsection{Age of the Universe}

The age of a flat universe is given by\footnote{There is an equivalent 
expression where the logarithm is replaced by $\sinh^{-1} \left( 
\sqrt{(1-\Omega_0)/\Omega_0} \, \right)$.}
\begin{equation}
H_0 t = \frac{2}{3} \, \frac{1}{\sqrt{1-\Omega_0}} \, \ln \left[ \frac{1+
	\sqrt{1-\Omega_0}}{\sqrt{\Omega_0}} \right] \,.
\end{equation}
At fixed $H_0$, lowering $\Omega_0$ increases the age. However, as already 
noted for the case of an open universe \cite{LLRV}, observations such as the 
galaxy correlation function more or less fix the combination $\Omega_0 H_0$, 
rather than $H_0$ itself. With that combination fixed, lowering $\Omega_0$ 
actually leads to a reduction of the age of the model universe (though a 
less severe one than in the open case). Consequently, $\Lambda$CDM models 
that fit other observations are {\em younger} the lower the density.

Observational age determinations remain controversial, perhaps solely 
because the older estimates are in conflict with almost any cosmology (and 
most seriously with {\em low}-density cosmologies, as seen in our figures). 
However, a recent analysis by Chaboyer et al. \shortcite{CKKD} (see 
also Bolte \& Hogan 1995; Jiminez et al. 1996) suggests a 95 
per cent lower limit of 12 Gyr, which is weak enough to leave most 
possibilities intact.

\subsection{Baryons in clusters}

The piece of evidence which is presently providing the strongest push
for researchers to consider low-density models is the question of the
baryon density in clusters \cite{WF,WNEF}. Typical quoted values suggest 
that about 20 per cent of material in clusters is in the form of baryons, 
primarily as hot intracluster gas. Taken at face value, this is consistent 
with standard nucleosynthesis only if the total density is substantially 
less than critical.

Although a pressing concern, there remain some potential problems with
the interpretation of the observations (see for example the summary by 
Steigman \& Felten 1995). A standard argument is that clusters are such 
large objects that the fractional baryon density within them must be 
representative of the Universe as a whole; however, individual measurements 
of the fractional baryon density are not all consistent with each other on a 
cluster-by-cluster basis (e.g. White \& Fabian 1995), so unless the 
observational errors have been somewhat underestimated this assumption 
must fail at some level. Nevertheless, this doesn't seem to be too important 
an uncertainty. For example, an unreasonably crude method of bringing the 
White \& Fabian \shortcite{WFab} results into agreement with one another is 
to throw out the clusters with the highest four or five determinations; even 
this only reduces the best-fitting baryon density by about 10 per cent. 
Retaining all of their preferred sample of 13 clusters, White \& Fabian find
\begin{equation}
\frac{\Omega_{{\rm B}}}{\Omega_0} = 0.05_{-0.015}^{+0.03} \, h^{-3/2} \,,
\end{equation}
at the 95 per cent confidence level. A further question concerns whether
or not the mass determinations (of either the gas or the total mass) are 
sufficiently sophisticated: for example, in a recent paper Gunn \& Thomas 
\shortcite{GT} carried out an analysis modelling the cluster gas as a 
multiphase medium, and found that the inferred fractional
baryon density can be reduced by a quarter or even a half. Finally, the
status of standard nucleosynthesis theory has been less static recently than 
for many years, with the upper limit on the baryon density rising 
substantially. Copi et al. (1995a,b) advocate the central value 
$\Omega_{{\rm B}} h^2 = 0.016$ that we are adopting, and their 95 per cent 
confidence upper limit is 0.024 (see also Kernan \& Krauss 1994; Sasselov \& 
Goldwirth 1995; Krauss \& Kernan 1995; Hata et al. 1995, and for an opposite 
opinion see Hogan 1995). Some recent deuterium abundance measurements in 
quasar absorption systems suggest even higher baryon fractions 
\cite{TFB,BT}.

Ignoring mass determination uncertainties beyond those accounted for by 
White \& Fabian \shortcite{WFab}, in combination the cluster and 
nucleosynthesis limits imply a central value $\Omega_0 = 0.32 h^{-1/2}$. 
Taken at face value the error bars exclude the critical-density case for any 
reasonable $h$ (the 95 per cent upper limit being $0.53 h^{-1/2}$), but in 
our view, shared by Babul \& Katz \shortcite{BK} and Gunn \& Thomas 
\shortcite{GT} (see also Balland \& Blanchard 1995), the theoretical 
uncertainties remain great enough that it would be extremely premature to 
abandon critical-density models completely. It is clear however that present 
interpretation of this vital constraint does favour values of $\Omega_0$ in 
the range between 0.2 and 0.6.

\subsection{Velocity flows}

\label{POT2}

A direct analysis of the velocity flow data has been carried out by Nusser 
\& Dekel \shortcite{ND} and Bernardeau et al. \shortcite{BJBD}. By analysing 
statistics of the velocity field in the weakly non-linear regime, they 
conclude that $\Omega_0$ must exceed 0.3 at the 95 per cent confidence 
level. The same result has also been obtained from an analysis of 
velocity outflows in void regions \cite{DR}. Summaries of these results are 
given in reviews by Dekel \shortcite{Dekel} and Strauss \& Willick 
\shortcite{SW}. The parameter region excluded by this constraint has 
only minimal overlap with our allowed regions, and always in a region where 
the age of the universe is anyway suspiciously low. So our analysis 
independently supports the conclusion of these papers. 

The results of those methods are quoted independently of any assumptions 
about the underlying power spectrum (though the former method does have some 
dependence on the assumption made). It would be expected that if the 
analyses were repeated with specific power spectra input, then the 
constraints would strengthen\footnote{After our paper was submitted,  
analyses based on computation of the power spectrum of mass fluctuations 
directly from the POTENT Mark III catalogue were produced \cite{KD,ZDHK}, 
including comparison with a range of cosmological models. Where they 
overlap, our results seem in good agreement with theirs.}.

\subsection{Lensing limits on $\Lambda$}

An extremely comprehensive account of lensing limits on the cosmological 
constant has recently been presented by Kochanek \shortcite{K95}. He 
concludes that the 95 per cent confidence limit for flat universes is 
$\Omega_0 > 0.34$, a constraint which is similar to that obtained from 
velocities. However, he notes that this limit is sensitive to several 
possible systematic biases, and it can be relaxed to some extent by the 
introduction of extinction due to dust obscuration in the lensing population 
of mainly E/S0 galaxies. 

\subsection{Type Ia supernovae}

Prospects now look extremely good for applying a classic cosmological test, 
using type Ia supernovae as standard candles at very large distances in 
order to study the deceleration of the Universe. Preliminary results from 
the Supernova Cosmology Project \cite{Petal} favour a decelerating Universe 
rather than an accelerating one, which if confirmed will give a stronger 
constraint than the lensing limits of the previous subsection. A 
decelerating flat universe requires $\Omega_0 > 2/3$. With many more 
supernovae already being analysed, this is perhaps the most promising route 
of all towards constraining low-density flat universes.

\subsection{Galaxy counts}

Faint galaxy counts in principle offer a fairly direct probe of the geometry 
of the Universe to high redshifts, but in practice the separation of 
geometry from evolution of the galaxy population has proven difficult. A 
recent analysis using {\it HST} observations \cite{DWPB} concludes with some 
definiteness that for a flat universe $\Omega_0$ must exceed 0.2, which is 
weaker than other constraints. A similar constraint had already been 
obtained from deep number counts by Gardner, Cowie \& Wainscoat 
\shortcite{GCW}. However, Driver et al. \shortcite{DWPB} add that with 
further assumptions concerning correctness of current morphological 
classifications, that limit can already be dramatically strengthened to 
$\Omega_0 > 0.8$ for a flat geometry. Further development of this constraint 
is clearly a promising avenue.

\section{Results}

\begin{figure*}
\centering
\leavevmode\epsfysize=11.3cm \epsfbox{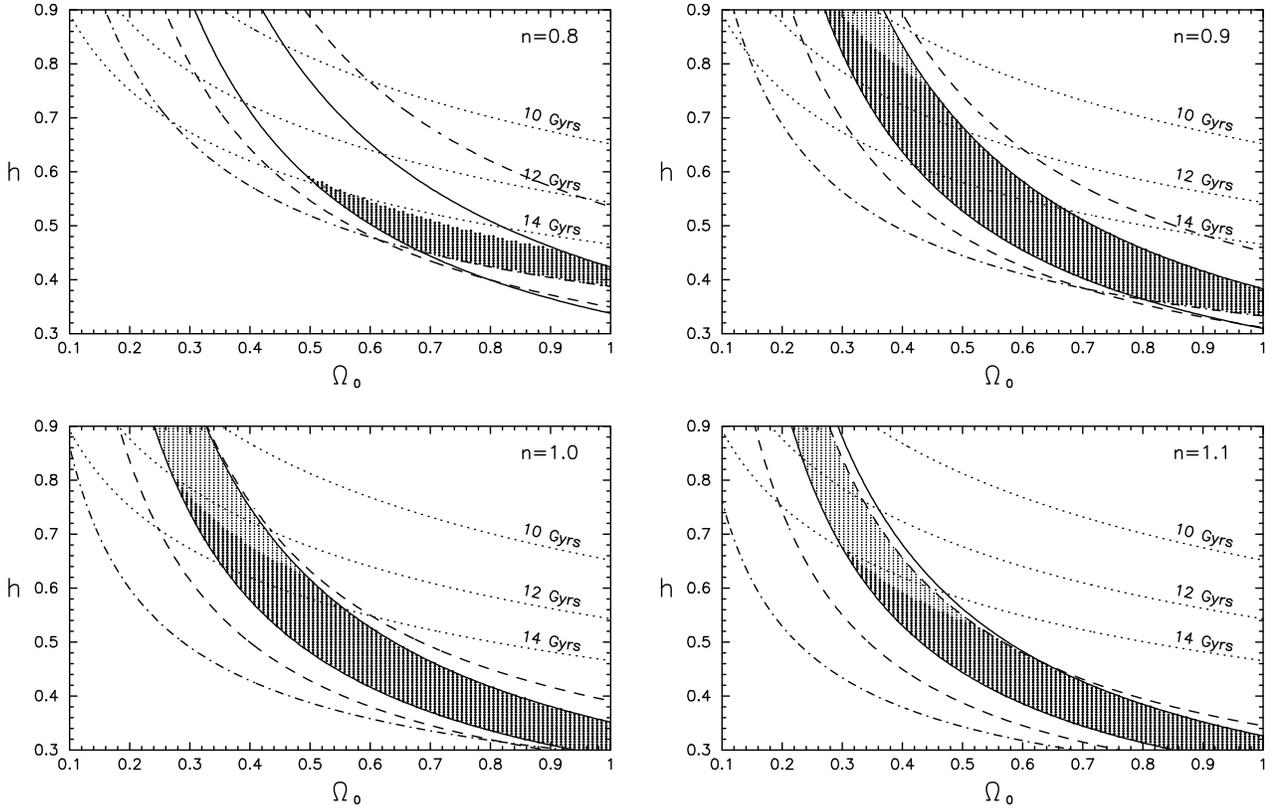}
\caption[figure1]{The constraints plotted in the $\Omega_0$--$h$ plane, for 
different $n$ in the case without gravitational waves. All constraints 
are plotted at 95 per cent confidence. The models are normalized to the {\it 
COBE} data. The solid lines are limits from the shape parameter, the dashed 
lines from the cluster abundance and the dot-dashed lines from the 
abundance of damped Lyman alpha systems. The POTENT constraint is not shown 
as it is weaker than all the other constraints. Contours of constant age are 
shown as dotted lines. The allowed region is shown with two different 
shadings, both highlighting the parameter space not excluded at more than 95 
per cent confidence on any single piece of data. The lighter shading shows 
models where the optical galaxies have to be anti-biased at $8 \, h^{-1}$ 
Mpc. The unshaded region in the $n=0.8$ plot which is allowed by all plotted 
data is excluded by Doppler peak height.}
\end{figure*}
 
Our results are shown in Figs 1 and 2, which indicate the microwave 
background and large-scale structure constraints, along with some age 
contours for reference. Let us first discuss Fig.~1, which shows the 
situation without gravitational waves.

Fig.~1 shows that incorporating the possibility of tilt greatly increases 
the space of viable parameters. By far the most important constraint is the 
shape of the galaxy correlation function; this constraint is reinforced by 
the cluster abundance constraint which lies in a similar region but is 
typically weaker. For low $n$, the allowed region is then further trimmed by 
the damped Lyman alpha system abundance. The POTENT constraint is weaker 
than the others in all parts of parameter space and we omit it for clarity. 
At low $\Omega_0$, models start to become dangerously young.

One can go to very high $n$ in these models. Although we haven't indicated 
it in a plot, we have found that there remains an allowed region even at $n 
= 1.4$. However, once $n > 1.3$, a strong anti-bias is necessary unless $h$ 
is well below 0.5. Also, such models predict a huge Doppler peak, and 
require a redshift of reionization which is just right to reduce the large 
peak down to the standard size.

In almost all regions of parameter space the height of the Doppler peak is 
not constraining. The exception is when $n = 0.8$, where it begins, quite 
dramatically, to cut off the high-$h$ region. To avoid confusion through 
plotting yet another line, we have indicated this simply by not shading the 
region (centred around $\Omega_0 = 0.5$, $h = 0.75$) that is allowed by 
other constraints but not the Doppler peak height. The exclusion is because 
for these parameters the peak is not high enough to fit the observations.
As $n$ is decreased, the Doppler peak constraint rapidly becomes very 
strong; already by $n = 0.75$ there is no allowed region at all. Note though 
that the value of $n$ at which this happens would be changed were the baryon 
density allowed to be substantially higher than indicated by 
nucleosynthesis.

The main problem with the Doppler peak constraint is that the upper limit 
cannot be used due to the possibility that reionization might bring the 
peak back down to an acceptable height. However, some regions of otherwise 
permitted parameters do require specific assumptions about reionization and 
we will discuss this shortly.

Except in rather extreme circumstances, the region allowed by these 
large-scale structure constraints is not further reduced by including the 
miscellaneous constraints from bulk velocity flows, gravitational lensing 
and galaxy counts that we discussed in Section 5, which favour $\Omega_0 > 
0.3$. Rather, our results provide independent support to the conclusions of 
those analyses. Type Ia supernovae \cite{Petal} seem the most promising 
route to a stronger constraint in the immediate future.

In Fig.~1, we have shaded the allowed regions in two different styles, to 
indicate whether anti-biasing of optical galaxies at $8 \, h^{-1}$ Mpc is 
required or not. In doing this, we take the {\it COBE} normalization to its 
95 per cent lower limit, so the light shaded regions definitely require 
anti-biasing. Whether this is excluded on any particular observational 
grounds is unclear. Motivated by Klypin et al. \shortcite{KPH}, we have also 
tried to calculate to what extent anti-bias is required at scales presently 
in the non-linear regime. We choose to consider the APM power spectrum 
measurement \cite{BE} at $1 \, h^{-1}$ Mpc. Using analytic approximations to 
estimate the non-linear power spectrum \cite{HKLM,PD,JMW,PD2} for the models 
we are interested in, we can determine which require an anti-bias at this 
scale. As it turns out, the resulting constraint is not much stronger than 
assuming no anti-bias at $8 \, h^{-1}$ Mpc using the linear power spectrum. 
In any case it seems less clear that anti-biasing is physically forbidden so 
far into the non-linear regime. The question of whether anti-bias is 
permitted will perhaps be accessible to new hydrodynamical $N$-body 
simulations, which might indicate whether or not forbidding anti-bias at $8 
\, h^{-1}$ Mpc can really be adopted as a firm constraint\footnote{Also see 
Kauffmann, Nusser \& Steinmetz \shortcite{KNS} for a recent discussion of 
the physical origin of biasing; their analysis does not favour 
anti-biasing.}.

Our conclusion in the case of no gravitational waves is that the large-scale 
structure data alone permit $\Omega_0$ as low as about 0.30, and show no 
preference for any particular value above that. However, at critical density 
the required Hubble parameter is very low (though see the more extensive 
discussion in Liddle et al. \shortcite{LLSSV} which also considers adding a 
hot dark matter component). The $\Lambda$CDM model is compatible with 
observations for a wide range of values of the tilt.

\begin{figure*}
\centering
\leavevmode\epsfysize=5.84cm \epsfbox{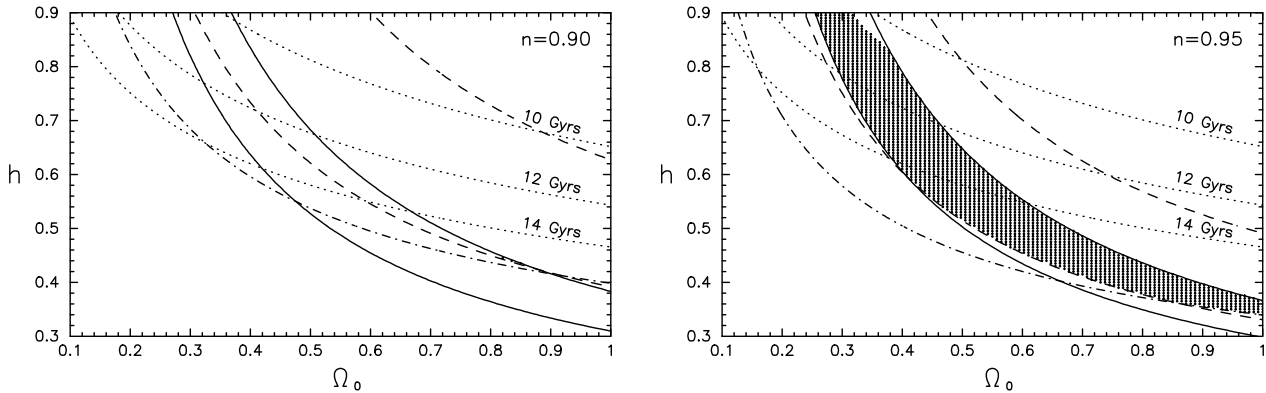}
\caption[figure2]{As Fig.~1, but with gravitational waves as given by 
power-law inflation. In each case where only a single line appears, the 
allowed region is to the upper right. In both of these plots, the unshaded 
region which is allowed by all plotted data is excluded by Doppler peak 
height.}
\end{figure*}

Fig.~2 shows what happens if gravitational waves are introduced in 
accordance with the power-law inflation model. This model requires $n<1$ so 
we only show two values. The introduction of gravitational waves leads to a 
much smaller allowed region than in their absence, with $n=0.9$ 
the smallest allowed value. At $n=0.9$, the shape and cluster constraints 
leave only a very narrow strip, which is then totally excluded by 
the Doppler peak constraint. For $n = 0.95$ there is an allowed region. The 
limit $n > 0.9$ in the power-law inflation model is stronger than that given 
by Liddle et al. \shortcite{LLSSV} in the critical-density case, because 
here we consider the additional constraint coming from the Doppler peak.

Although we will not fully investigate the issue here, the Doppler peak 
constraint does have some implications for reionization. Because 
$\Lambda$CDM models at low density have a very high normalization 
(especially at high redshift), early reionization is more likely in these 
models than in other cosmologies \cite{LL95}. However, the physical 
understanding of the reionization process remains poor enough that most 
estimates of the reionization redshift can be interpreted only as upper 
limits.

We'll restrict ourselves to the case with no gravitational waves, and for 
simplicity assume the central nucleosynthesis value for $\Omega_{{\rm B}}$ 
(to which the results are quite sensitive). In some regions of parameter 
space, any early reionization would remove the Doppler peak sufficiently to 
be incompatible with observation, while in others early reionization is 
necessary to bring it down to acceptable levels. Using the formulae given 
earlier, we find the following results. For $n \geq 1.1$, 
reionization at any $z_{{\rm R}}$ less than 15 is not enough to allow any 
interesting models. For lower $n$, reionization becomes less necessary in 
order to lower the Doppler peak; indeed, it becomes a worry that it may 
lower it too much. If $n = 0.8$, then $z_{{\rm R}} \geq 15$ is sufficient to 
exclude all the otherwise allowed region. A full examination of the 
relationship between large-scale structure constraints and the required 
reionization properties, taking into account the uncertainty in the baryon 
density, is clearly warranted.

\section{Conclusions}

Our results extend previous work both in terms of the range of parameter 
space studied and in terms of the observations considered. By including the 
possibility of tilt and gravitational waves, we have explored the 
parameter space available for inflationary-based $\Lambda$CDM models. Recent 
work (Krauss \& Turner 1995; Ostriker \& Steinhardt 1995; Bagla et al. 1995) 
has favoured flat low-density models primarily not for large-scale 
structure reasons, but rather as a resolution of two separate conflicts, the 
first being age verses the Hubble constant and the second being the cluster 
baryon fraction verses nucleosynthesis. We have shown that while 
compatibility with large-scale structure does not in itself select favoured 
values of $\Omega_0$, it is compatible with low-density $\Lambda$CDM models.

It is interesting to compare these results with the open CDM case explored 
using the same techniques in Liddle et al. \shortcite{LLRV}. The principal 
difference is that at low $\Omega_0$ the {\it COBE} normalization is very 
different, so that in the open case anti-bias is not necessary. 
However, other observations, such as the cluster abundance, shift as well to 
maintain a substantial allowed region. The allowed regions are fairly 
similar in the $\Lambda$CDM and open CDM cases, with no particular 
preference for either on present data, though the open models tend to prefer 
a slightly higher value of $n$. In going to a $\Lambda$CDM model, there is 
only a very modest gain in age relative to an open CDM model, so the 
principal motivation for introducing the cosmological constant is in 
maintaining consistency with standard inflationary models rather than 
because of an age crisis.

Returning to our discussion of the $\Lambda$CDM model, an extremely useful 
result is the fitting function, equation (\ref{norm}), which allows a quick 
normalization of all $\Lambda$ models, since to an excellent approximation 
it is independent of the parameters $h$ and $\Omega_{{\rm B}}$ and the 
presence of any component of hot dark matter. We have also made the first 
detailed comparison of these models with the observational situation 
concerning intermediate-scale CMB anisotropies; though observations there 
are still at a very primitive level compared with what one hopes will become 
possible in the future, already one can obtain new constraints against 
regions of parameter space involving substantial gravitational waves, which 
are viable against all other present observations. We have also noted that 
certain areas of parameter space are viable against this constraint only if 
reionization occurred suitably; in some regions it is required to happen 
fairly early, while in others it is forbidden from doing so.

In the case of a scale-invariant spectrum with no gravitational waves, our 
results are in broad agreement with those obtained by other authors. The 
main challenge from large-scale structure for the low-density versions of 
these models continues to be the question of whether it is reasonable to 
believe that optical galaxies are un- or even anti-biased. We have included 
more constraints than the earlier treatments but for scale-invariant spectra 
they are not more constraining.

By extending the parameter space to include tilt and gravitational waves, we 
have found a large region compatible with present data. Amongst low-density 
models, it seems $\Omega_0$ in the range 0.4 to 0.5 is better than
$\Omega_0=0.3$ (continuing a historical upward trend in the preferred
`low-density' $\Omega_0$), as is also the case for open CDM models 
\cite{LLRV}. However, large-scale structure studies do not look particularly 
promising for lifting the degeneracy of allowed $\Omega_0$ values; the types 
of evidence we discussed in Section 5 appear more promising for delivering 
an ultimate verdict for or against the cosmological constant. 

\section*{ACKNOWLEDGMENTS}

ARL is supported by the Royal Society, PTPV by the PRAXIS XXI programme of 
JNICT (Portugal), and MW by DOE. ARL, DHL and MW thank the Aspen Center for 
Physics where discussions which led to this paper took place, and ARL thanks
TAC (Copenhagen) for hospitality while some of the work was carried out. MW 
thanks Avery Meiksin for conversations. PTPV acknowledges the use of the 
Starlink computer system at the University of Sussex. 


\bsp
\end{document}